\documentclass[12pt]{article}
\usepackage{amssymb}

\newcommand{\be}{\begin{equation}}
\newcommand{\ee}{\end{equation}}
\newcommand{\bea}{\begin{eqnarray}}
\newcommand{\eea}{\end{eqnarray}}

\title{Radiaitve Breaking of Conformal Symmetry \\ in the Standard Model}

\author{A.B. Arbuzov$^{1,2}$,
R.G. Nazmitdinov$^{1,3}$,
A.E. Pavlov$^{4}$, \\
\fbox{V.N. Pervushin$^{1}$}, 
A.F. Zakharov$^{1,5}$}

\date{}

\begin{document}

\maketitle

\begin{center}
$^{1}$Bogoliubov Laboratory for Theoretical Physics,
        Joint Institute of Nuclear Research, Dubna, 141980, Russia \\
$^{2}$Department of Higher Mathematics, Dubna State University,
        Dubna, 141980, Russia \\
$^{3}$Department de F{\'\i}sica, Universitat de les Illes Balears,
        Palma de Mallorca, E-07122, Spain \\
$^{4}$Moscow State Agri-Engineering University, Moscow, 127550, Russia \\
$^{5}$Institute of Theoretical and Experimental Physics, Moscow, 117259, Russia
\end{center}


\begin{abstract}
Radiaitve mechanism of conformal symmetry breaking
in a comformal-invariant version of the Standard Model
is considered. The Coleman-Weinberg mechanism of dimensional
transmutation in this system gives rise to finite vacuum
expectation values and, consequently, masses of scalar and
spinor fields. A natural bootstrap between the energy
scales of the top quark and Higgs boson is suggested.
\end{abstract}

\section{Introduction}

In spite of the recent discovery of the Higgs boson at LHC, we still do not
have clear answer about the fundamental origin(s) of the electroweak
energy scale.
The same concerns the question about the origin
of the QCD energy scale $\Lambda_{\mathrm QCD}$. There is a principal
difference in the treatment of these scales within the Standard Model
(SM)~\cite{Glashow:1961tr,Weinberg:1967tq,Salam:1968rm}.
Namely, the electroweak scale $(\sim 100)$~GeV is provided by the value
of the tachyon mass parameter in the primary SM Lagrangian, while the QCD scale
is not related to any parameter of the Lagrangian.
It is commonly assumed  that $\Lambda_{\mathrm QCD}$ appears due to the mechanism
of the dimensional transmutation caused by the conformal anomaly in the QCD
Lagrangian (and that could happen even with massless quarks).
In this letter we suggest a simple reduction of the electroweak
SM to its conformal-invariant version and show that the
mechanism of the dimensional transmutation gives rise to 
condensates and masses for scalar (the Higgs boson) and fermion
(the top quark) fields.

According to the general wisdom, all SM particles (may be except neutrinos)
own masses due to couplings with the Higgs boson vacuum expectation value.
This expectation value is brought about by the spontaneous breaking of the global symmetry
in the Higgs sector~\cite{Englert:1964et,Higgs:1964pj}.
In the SM, one deals with the potential
\be \label{5a}
V_{\rm Higgs}(\phi)=\frac{\lambda^2}{2}(\phi^\dagger\phi)^2 + \mu^2\phi^\dagger\phi,
\ee
where one component of the complex scalar doublet field
$\phi=\left(\begin{array}{c}\phi^+\\ \phi^0\end{array}\right)$
acquires a non-zero vacuum expectation value $\langle\phi^0 \rangle = v/\sqrt{2}$
if $\mu^2<0$ (the stability condition $\lambda^2>0$ is assumed).
The presence of the tachyon-like mass term in the potential is crucial for
this construction. In contrast to the spontaneous symmetry breaking,
it breaks the conformal symmetry {\em explicitly} being
the only {\em fundamental} dimensionful parameter in the SM.
We recall that the explicit breaking of the  conformal symmetry in the
Higgs sector gives rise to the serious problem of fine tuning (or naturalness)
in the renormalization of the Higgs boson mass,
that is certainly one of the most unpleasant features of the SM.

In the classical approximation, the condition of the potential minimum  yields
the relation between the vacuum expectation value and the primary parameters
$\mu$ and $\lambda$ in the form $v=\sqrt{-2{\mu^2}}\,/\lambda$.
Within the SM this quantity is related as well to the Fermi coupling constant,
derived from the muon life time measurements:
$v=(\sqrt{2}G_{\mathrm{Fermi}})^{-1/2}\approx 246.22$~GeV.

It was shown that the measured value of the Higgs boson mass makes the SM being
self-consistent up to very high energies of the order of the Planck mass
scale~\cite{Bezrukov:2012sa,Alekhin:2012py,Bednyakov:2015sca}.
Direct and indirect experimental searches push high up
the possible energy scale of
new physical phenomena. In this situation the question, why the top quark mass,
the Higgs boson mass, and the electroweak (EW) scale $v$ are of the same order,
becomes more and more intriguing.

The idea about dynamical breaking of the EW gauge symmetry
with the aid of the top quark condensate was continuously discussed in the literature since
the pioneering papers~\cite{Nambu:1989sx,Miransky:1988xi,Nambu:1990hj,Bardeen:1989ds}
(see the state of art, for example, in~\cite{Cvetic:1997eb,Volovik:2013bjq,Matsuzaki:2012mk}
and references therein).
It has been already mentioned by P.~Higgs~\cite{Higgs:1964pj} that a
EW symmetry-breaking scalar field can be described by
a non-elementary bilinear combination of Fermi fields.
It is especially noteworthy that such approaches require the introduction
of new interactions beyond the SM.
For example, it can be an interaction of the four-fermion type, similar to the one
in the Nambu--Jona-Lasinio model. However, numerous high-energy experiments, which
are continuously looking for signals of such new physics scenarios, do not manifest
any success in this direction.

We recall that quark condensates correspond to quadratically divergent tadpole
loop diagrams in the Quantum Field Theory. In the perturbative QCD
such diagrams have zero weights due to the gradient invariance of
the gluon-quark interaction. In fact, this can explain why there is no
any direct information on the top-quark condensate value from high-energy
observables\footnote{Obviously, due to the short life time of the top quark,
there is no chance to probe its condensate at low energies as well.}.
Nevertheless, in the same way as in the low-energy QCD, scalar degrees of freedom
can have non-zero contributions due to their effective interaction
with fermion condensates. As a matter of fact, in the SM there is a
coupling of a fundamental scalar to a fermion pair.
Within the SM, such single-propagator one-loop integral contributions are known to cancel
out everywhere except the Higgs sector, where they lead to the fine tuning problem
mentioned above. It is highly unlikely ({\it i.e.} only due to an extreme fine tuning)
that the proper renormalization will completely diminish such terms.
We underline that the very existence of quark condensates has nothing
to do with four-fermion interactions.

All mentioned facts suggest that:
i) it might be worth to examine the possibility
of a tight relation between the top quark and Higgs boson mass scales;
ii) further to clarify a relationship between a conformal anomaly
and the appearance of the EW energy scale.
As discussed in ref.~\cite{Bardeen:1995kv}, the radiative stability of
the Higgs boson mass, {\it i.e.}, a resolution of the naturalness problem,
can be ensured by the classical scale invariance.
Attempts to exploit the classical scale invariance in construction of the SM
became rather popular in the literature,
see, {\it e.g.},~\cite{Oda:2013uca,Khoze:2013uia,Farzinnia:2013pga,Gorbunov:2013dqa},
but most of them introduce specific new interactions beyond the SM.

\section{Radiative Breaking of Conformal Symmetry}

To begin with, we recall the mechanism of radiative breaking of the
scale invariance introduced by S.~Coleman and E.~Weinberg~\cite{Coleman:1973jx},
which we name the CW mechanism.
They have shown that the renormalization of several classical scale-invariant Lagrangians
leads to a spontaneous breaking of the scale invariance.  The key reason is that those
model possess an instability with respect to infrared singularities at the quantum level.
In particular, in the considered massless $\phi^4$ model a
stable minimum of the effective potential seems to lie outside the region of the
one-loop approximation applicability.

But it was shown also that such a mechanism works for a system of scalar and
vector fields with abelian or non-abelian interactions. It is commonly supposed that
the dimensional transmutation in the QCD~\cite{Collins:2011zzd} has the same
radiative symmetry breaking origin due to the conformal anomaly,
even so that the perturbative approximation does not work in the QCD at low
energies. Modern simulations of the QCD on lattice support this conjecture.
In the present paper we extend the Coleman-Weinberg formalism of the
radiative symmetry breaking for a system of scalar and
fermion fields with Yukawa interactions.

Let us start at the classical level with the conformal-invariant Lagrangian
describing one scalar and one fermion field:
\be \label{L_cl}
L_{\mathrm{cl}} = \frac{1}{2}(\partial_\mu\phi_c)^2 - \frac{\lambda^2}{2}\phi_c^4
+ i\bar{\Psi}_c\gamma_\mu\partial_\mu\Psi_c
- y \phi_c\,\bar{\Psi}_c\Psi_c,
\ee
where $\phi_c$ and $\Psi_c$ are classical massless fields.
This model is obviously renormalizable. At the quantum level
(index ``$c$'' is then removed)
we have to add counterterms of all possible kinds:
\begin{eqnarray} \label{L_ct}
L_{\mathrm{c.t.}} &=& \frac{1}{2}A(\partial_\mu\phi)^2
- \frac{1}{2}B\phi^2
- \frac{1}{2}C\phi^4
+ iD\bar{\Psi}\gamma_\mu\partial_\mu\Psi
- E\bar{\Psi}\Psi
+ G\phi\,\bar{\Psi}\Psi
+ F\phi.
\end{eqnarray}
The classical scale invariance provides the explicit condition
for the mass-like counterterms $B$ and $E$:
they should cancel out with the relevant loop diagram contributions.
Note that in the one-loop approximation
the term $F\phi$ corresponds to the tadpole fermion loop diagram, {\it i.e.},
$\phi\langle\bar\Psi\Psi\rangle$.
The latter is proportional to the integral
\be \label{Int}
\int\frac{d^4k}{i\pi^2}\frac{\mathrm{Tr}(k_\mu\gamma_\mu+m_f)}{k^2-m^2_f+i\varepsilon}
\ee
which is exactly zero for a massless fermion $(m_f=0)$.
However, according to ref.~\cite{Coleman:1973jx},
it is not possible to preserve the {\em classical scale invariance}
in a wide class of systems at the quantum level because of infrared
instabilities in quantum loop corrections.

So, it turns out that the
renormalization point should be shifted away from the origin $\phi=0$.
The so called {\em radiatively induced symmetry breaking} takes place
in the system because the effective potential of the scalar field
is infrared divergent at $\phi=0$.
We extend this idea to the model~(\ref{L_cl})
that at the quantum level should be renormalized
at $\phi=M\neq 0$, where $M$ is an energy scale.
In other words, the classical conformal-invariant theory
possesses a quantum instability, called {\em conformal anomaly}.
Nevertheless, the breaking of the symmetry is {\em dynamical}
or spontaneous\footnote{According to the classification of the types of symmetry breaking
suggested by Y.~Nambu~\cite{Nambu:1990dx}.}, and the classical symmetry
continues guiding the system.
As the result, the so called {\em dimensional transmutation} happens,
a dimensionless parameter of the classical Lagrangian can be traded
for a dimensionful one (actually for a ratio of some energy scales).
After shifting the renormalization point from $\phi=0$,
we have to perform the following steps:
1) look for the minimum of the effective potential;
2) analyze the masses of our fields in this point;
3) test the stability of the system in the resulting potential.
The latter condition should be the principal one in construction of
a physical model.

As was proved by Coleman and Weinberg~\cite{Coleman:1973jx},
even a pure $\phi^4$ self-interaction of a scalar field gives
rise to a radiative breaking of the conformal symmetry.
However, to
obtain a stable solution for the effective potential minimum
a coupling to a gauge field was added.
In our consideration we use a Yukawa coupling to a fermion field.
This field yields an additional contribution to the effective potential
and helps effectively to reach a stable minimum in the perturbative
domain of the coupling constant values.
As it is stated in ref.~\cite{Coleman:1973jx}
``{\em there is no obstacle to extending the formalism}'' for this case, and it does
not matter ``{\em whether we couple our external sources to fundamental or composite
fields}''. 

Let us study the fermion condensate $\langle\bar\Psi\Psi\rangle$. Its value
is proportional to the integral~(\ref{Int}). The latter is extremely unstable
with respect to appearance of a mass of the fermion. Direct calculations show that
even a tiny (but non-zero mass) makes this integral being quadratically divergent:
\be
\langle\bar\Psi\Psi\rangle \sim m_f\Lambda^2,
\ee
where $\Lambda$ is an ultraviolet cut-off.  
Then it should be renormalized by a corresponding counter term.
The classical conformal symmetry condition explicitly requires
a complete cancellation of this divergent loop contribution by the
corresponding counterterm $E\overline{\Psi}\Psi$. 
Similar to the consideration for a scalar field in ref.~\cite{Coleman:1973jx},
we are forced to shift the renormalization
point to a non-zero value. 

Therefore, it seems reasonable to conjecture that
our model possesses a loop-back
effect: 
i)if a non-zero value for the scalar field condensate appears (due to a quantum effect),
it immediately yields a mass for the fermion; 
ii) the fermion field creates also a non-zero
(even divergent) condensate; iii) the condensate allows to the scalar
field to have a nonzero vacuum expectation value.
Of course, this is just a schematic description of the loop-back effect,
while one has to find a stable self-consistent solution of the system as a whole.
Moreover, the proper renormalization should be applied consistently for the both fields.
Unfortunately, the system of equations for the effective potential even at one loop approximation
is rather non-linear, and we do not have exact results. In this sense our problems is
similar to the QCD one. Nevertheless, our conjecture is that the CW mechanism
works for our case, and there exists a stable solution.

As a next step, we consider the case of the conformal-invariant Lagrangian of
the Higgs boson interactions
\be \label{L_int}
L_{\mathrm{int}} = - \frac{\lambda^2}{2}(\Phi^\dagger\Phi)^2 - y_t \Phi \bar{t}t,
\ee
where only the most intensive terms are listed:
the self-interaction and the Yukawa one with the top quark.
Note that we have dropped the tachyon mass term from the SM.

Further, we assume that the $O(4)$ symmetry of the Higgs
sector should be spontaneously broken to the $O(3)$
symmetry\footnote{Strictly speaking in the Higgs sector, we have spontaneous breaking
of the global $SU(2)_L\times SU(2)_R$ symmetry down the the custodial $SU(2)_V$ symmetry.}.
This assumption will be confirmed below.
As a result, the construction should give rise to a non-zero
Higgs field vacuum expectation value $v$.

Vacuum averaging with subsequent renormalization of the fermion operators in
eq.~(\ref{L_int}) leads to the potential of the form
\bea\label{V_cond}
V(h) &=& \frac{\lambda^2}{8}h^4 + \frac{y_t}{\sqrt{2}}\langle \bar t\, t\rangle h.
\eea
The extremum condition for the potential $dV(h)/(dh)|_{h=v}=0$
yields the relation
\be \label{lambda}
v^3 \frac{\lambda^2}{2} = - \frac{y_t}{\sqrt{2}}\langle \bar t\, t\rangle.
\ee
The nontrivial solution of the minimum condition leads
to the standard decomposition $h = v  + H$, where $H$ represents excitations
(non-zeroth harmonics) with the condition $\int d^3x H=0$. The Yukawa coupling of
the top quark $y_t \approx 0.99$ is known from the experimental value
of the top quark mass $m_t=v y_t/\sqrt{2} \simeq 173.2$~GeV~\cite{Agashe:2014kda}.
Thus, the spontaneous symmetry breaking yields the potential minimum which results
in the non-zero vacuum expectation value $v$ and Higgs boson mass.
In fact, the substitution $h = v  + H$ into the potential~(\ref{V_cond}) gives
\be
V_{\rm cond}(h) = V_{\rm cond}(v) + \frac{m_H^2}{2}H^2
+ \frac{\lambda^2 v}{2} H^3+\frac{\lambda^2}{8}H^4,
\ee
which defines the scalar particle mass as
\be
\label{mh}
m_H^2\equiv\frac{3\lambda^2}{2}v^2.
\ee
We stress that this relation is different from
the one $(m_H=\lambda v)$ that emerges in the SM with the standard
Higgs potential~(\ref{5a}).

With the aid of eqs.~(\ref{lambda}) and (\ref{mh}),
the squared scalar particle mass can be expressed in terms of the top
quark condensate:
\be \label{h_mass}
m^2_H = -\frac{3 y_t \langle \bar t\, t\rangle}{v \sqrt{2}}.
\ee
To have $m_H=125$~GeV we need
\be \label{tt_value}
\langle \bar t\, t\rangle \approx -(122\ {\mathrm{GeV}})^3.
\ee
As discussed above, such a value of the top quark condensate does not
affect the low energy QCD phenomenology.

The value of the top quark condensate in our case should correspond
to a certain adjustment of the divergent loop integral renormalization.
This adjustment resembles a fine-tuning. However, it is quite different
from the fine-tuning in the Higgs boson mass renormalization. 
First, note that
the energy scale of the top quark condensate appears to be the same as the
general electroweak one given by the Higgs boson mass and VEV values.
Considering the explicit examples of the radiative symmetry breaking~\cite{Coleman:1973jx},
we find that the generated values of masses and condensates are defined by
the values of the renormalization scale and the coupling constants.
In fact, we substitute the tachyon mass by a single renormalization scale
of the CW mechanism. This renormalization scale is not an "additional" one
to the SM, in our scenario it is the only second one after $\Lambda_{\mathrm{QCD}}$.
The value of this scale can not be derived starting from the conformal Lagrangian.
Therefore, we adjust the renormalization condition to an observable in eq.~(\ref{tt_value}).

Let us analyse also a similar structure observed in the low-energy QCD.
It is natural assume that the scale of the light quark condensate,
$\langle \bar u\, u\rangle \approx -(250\ {\mathrm{MeV}})^3$,
is related to the scale of the conformal anomaly in the QCD. At the same time,
those anomalous properties of the QCD vacuum lead to the constituent
mass of a light quark to be of the order $300$~MeV.
Some anomalous properties of the {\em relevant} vacuum
give rise to the mass of top quark\footnote{Certainly, QCD effects both
in the mass and in the condensate value of the top quark are relatively small
compared to the Yukawa ones.} and to the condensate being of the same energy scale.

Although we have dropped the scalar field mass term from the
classical Lagrangian, it will re-appear after quantization and subsequent renormalization.
In fact, such a counter-term in the Higgs sector is necessary.
According to ref.~\cite{Bardeen:1995kv}, the conformal symmetry of the classical
Lagrangian will lead just to  the proper quantity in the mass term being consistent
with all other quantum effects.
A similar situation takes place in QCD: the chiral symmetry at the quark level
re-appears at the hadronic level even so that the breaking is obvious~\cite{Witten:1983tw}.

\section{Conclusions}

We suggest  the mechanism of radiative breaking of the conformal
symmetry in the Standard Model.
This enables us to resolve the problem of the regularization of quadratically divergent
tadpole loop integrals by relating them to the condensate values extracted from
the experimental observations.
In our construction, the top quark condensate supersedes the tachyon-like mass
term in the Higgs potential.
The considered mechanism allows to establish relations between condensates
and masses including the Higgs boson one.
In this way, we propose a simple bootstrap between the Higgs and top
fields (and their condensates).

Our approach is similar to the one commonly accepted in the QCD.
In fact, the conformal symmetry breaking in the QCD provides a single energy scale
 for the light-quark and gluon condensates as well as for the constituent quark mass.
For the time being, we are not able to describe these phenomena in the QCD using only
its Lagrangian.
We can, however, extract the relevant scales from observables.
In the same manner, the scale of the top quark mass, which appears due to
the Higgs condensate, might be naturally related to the scale of its own condensate.

It is noteworthy that we consider the Higgs boson as an elementary particle,
without introduction of any additional interaction beyond the SM.
After the spontaneous symmetry breaking in the tree-level
Lagrangian, the difference from the SM appears only in the value of the
Higgs boson self-coupling constant. The latter can be extracted from the
 LHC data only after the high-luminosity upgrade,
and it will be certainly measured at a future linear $e^+e^-$ collider.

{\bf Acknowledgments.}
The authors are grateful to  Yu.~Budagov, A.~Efremov and N.~Kochelev
for useful discussions. ABA acknowledge the Dynasty foundation for support.
VNP was supported in part by the Bogoliubov--Infeld program. 
AEP and AFZ are grateful to the JINR Directorate for hospitality.


\begin{thebibliography}{99}

\bibitem{Glashow:1961tr}
  S.~L.~Glashow,
  Nucl.\ Phys.\  {\bf 22} (1961) 579.

\bibitem{Weinberg:1967tq}
  S.~Weinberg,
  Phys.\ Rev.\ Lett.\  {\bf 19} (1967) 1264.

\bibitem{Salam:1968rm}
  A.~Salam,
  Conf.\ Proc.\ C {\bf 680519} (1968) 367.

\bibitem{Englert:1964et}
  F.~Englert and R.~Brout,
  Phys.\ Rev.\ Lett.\  {\bf 13} (1964) 321.

\bibitem{Higgs:1964pj}
  P.~W.~Higgs,
  Phys.\ Rev.\ Lett.\  {\bf 13} (1964) 508.

\bibitem{Bezrukov:2012sa}
  F.~Bezrukov, M.~Y.~Kalmykov, B.~A.~Kniehl and M.~Shaposhnikov,
  JHEP {\bf 1210} (2012) 140.

\bibitem{Alekhin:2012py}
  S.~Alekhin, A.~Djouadi and S.~Moch,
  Phys.\ Lett.\ B {\bf 716} (2012) 214.

\bibitem{Bednyakov:2015sca}
  A.~V.~Bednyakov, B.~A.~Kniehl, A.~F.~Pikelner and O.~L.~Veretin,
  Phys.\ Rev.\ Lett.\  {\bf 115} (2015) no.20,  201802.

\bibitem{Nambu:1989sx}
  Y.~Nambu,
  ``Model building based on bootstrap symmetry breaking,
  Published in Proc. 1989 Workshop on Dynamical Symmetry Breaking,
  eds. T.~Muta and K.~Yamawaki, Nagoya University, Nagoya, 1990.

\bibitem{Miransky:1988xi}
  V.~A.~Miransky, M.~Tanabashi and K.~Yamawaki,
  Phys.\ Lett.\ B {\bf 221} (1989) 177.

\bibitem{Nambu:1990hj}
  Y.~Nambu,
  ``New bootstrap and the standard model'',
  Published in Proceedings. Dalitz conference. Plots, quarks and strange particles,
  Oxford University Press, Oxford, eds. I.J.R.~Aitchison, C.H.~Llewellyn Smith, J.E.~Paton,
  1990, pp.56-65.

\bibitem{Bardeen:1989ds}
  W.~A.~Bardeen, C.~T.~Hill and M.~Lindner,
  Phys.\ Rev.\ D {\bf 41} (1990) 1647.

\bibitem{Cvetic:1997eb}
  G.~Cvetic,
  Rev.\ Mod.\ Phys.\  {\bf 71} (1999) 513.

\bibitem{Volovik:2013bjq}
  G.~E.~Volovik and M.~A.~Zubkov,
  JETP Lett.\  {\bf 97} (2013) 301
   [Pisma Zh.\ Eksp.\ Teor.\ Fiz.\  {\bf 97} (2013) no.6,  344].

\bibitem{Matsuzaki:2012mk}
  S.~Matsuzaki and K.~Yamawaki,
  Phys.\ Lett.\ B {\bf 719} (2013) 378.

\bibitem{Bardeen:1995kv}
  W.~A.~Bardeen,
  ``On naturalness in the standard model,''
  FERMILAB-CONF-95-391-T, C95-08-27.3.


\bibitem{Oda:2013uca}
  I.~Oda,
  Adv.\ Stud.\ Theor.\ Phys.\  {\bf 8} (2014) 215.

\bibitem{Khoze:2013uia}
  V.~V.~Khoze,
  JHEP {\bf 1311} (2013) 215.

\bibitem{Farzinnia:2013pga}
  A.~Farzinnia, H.~J.~He and J.~Ren,
  Phys.\ Lett.\ B {\bf 727} (2013) 141.

\bibitem{Gorbunov:2013dqa}
  D.~Gorbunov and A.~Tokareva,
  Phys.\ Lett.\ B {\bf 739} (2014) 50.

\bibitem{Coleman:1973jx}
  S.~R.~Coleman and E.~J.~Weinberg,
  Phys.\ Rev.\ D {\bf 7} (1973) 1888.

\bibitem{Collins:2011zzd}
  J.~Collins,
  ``Foundations of perturbative QCD,''
  Cambridge monographs on particle physics, nuclear physics and cosmology, 2011,
  pp. 57-58.

\bibitem{Nambu:1990dx}
  Y.~Nambu,
  ``Dynamical symmetry breaking,''
  In Tokyo 1990, Evolutionary trends in the physical sciences, pp. 51-66.

\bibitem{Agashe:2014kda}
  K.~A.~Olive {\it et al.} [Particle Data Group],
  Chin.\ Phys.\ C {\bf 38} (2014) 090001.

\bibitem{Witten:1983tw}
  E.~Witten,
  Nucl.\ Phys.\ B {\bf 223} (1983) 422.

\end{thebibliography}
\end{document}